# A HYBRID DESIGN OF PROJECT-X*

W. Chou[#], Fermilab, Batavia, IL 60510, U.S.A.

*Abstract*

Project-X is a leading candidate of the next major accelerator construction project at Fermilab. The mission need of Project-X is to establish an intensity frontier for particle physics research, or more precisely, to build a multi-MW proton source for neutrino and other particle studies. Coupled with an upgraded Main Injector (MI) and Recycler, an 8 GeV superconducting RF (SRF) H⁻ linac meets this need [1]. However, a more cost effective approach would be a hybrid design, namely, a combination of a 2 GeV SRF linac and an 8 GeV rapid cycling synchrotron (RCS) in lieu of an 8 GeV SRF linac. This alternative design also meets the mission need but at a lower cost since a synchrotron is cheaper than a SRF linac. It retains the ability to use a 2 GeV SRF linac for ILC technology development. It reuses the existing Debuncher enclosure and Booster RF. The transport line of 2 GeV H⁻ particles is shorter than the present 8 GeV design since stronger bending magnets can be used. The blackbody radiation stripping of H⁻ particles will no longer be a problem and the requirement of a cryogenic beam screen can be eliminated. The efficiency of stripping foil is higher and injection loss (kJ) will be reduced by a factor of 4. This paper introduces this alternative design and describes briefly the major components in the design.

## INTRODUCTION

According to the DOE Order 413.3A, "*The Conceptual Design process requires a mission need as an input. Concepts for meeting the need are explored and alternatives considered to arrive at a set of alternatives that are technically viable, affordable and sustainable.*" (Attachment 3, Page 1)

Fermilab is currently pursuing future facilities on both the energy and intensity frontiers. The leading candidate is a multi-MW proton source nicknamed Project-X. The mission need of Project-X is to bring the U.S. to the intensity frontier of particle physics research. More precisely, the goal is to build a facility that would provide ~2 MW particle beams at 120 GeV and a fraction of MW at 8 GeV. The present design calls for an 8 GeV superconducting RF (SRF) H⁻ linac as a principal component of this new facility [1]. It can meet the mission need. However, it is by no means the only design. There was an early design based on an 8 GeV rapid cycling synchrotron (RCS) [2]. Each design has its pros and cons. This paper proposes a third approach – a hybrid design. It consists of a 2 GeV SRF H⁻ linac and a 2-8 GeV RCS. It presents a number of attractive features. In particular, the construction cost could be significantly lower than an 8 GeV SRF linac. The cost saving comes from the following facts:

- The cost of a 2-8 GeV RCS is lower than a 2-8 GeV SRF linac.
- A transport line of 2 GeV H⁻ particles is shorter and simpler than an 8 GeV one, because it allows the use of stronger bends and eliminates the need for a cryogenic beam screen.
- The RCS can reuse the existing Debuncher enclosure.
- The RCS can reuse the existing Booster RF system.

Furthermore, this hybrid design provides a flexible configuration. Namely, if necessary, one can stage the linac construction, e.g. Phase I – 600 MeV, Phase II – 2 GeV. Note that a stand-alone linac below 8 GeV without an RCS cannot be connected to the Main Injector because of the limited tuning range of the MI RF system.

In addition to cost saving and flexibility, the hybrid design presents several other advantages:

- A 2 GeV SRF linac retains the ability to serve as an ILC SRF technology development project.
- Transportation of H⁻ ions is easier at 2 GeV than at 8 GeV because magnetic stripping is weaker and blackbody radiation stripping is not an issue.
- Conversion of H⁻ to H⁺ ions by stripping foils is more efficient at 2 GeV than at 8 GeV because the interaction cross section is larger at lower energy.
- The injection loss (a major concern in high intensity operation) in terms of kJ is 4 times smaller at 2 GeV than at 8 GeV.
- This design can piggy-back on the present NOvA ANU Project, which will use the Recycler as a proton accumulator for eliminating the injection front porch in the Main Injector.
- Fermilab has four decades of experience in synchrotrons and a world-class army in this highly specialized technical field. This design provides a way to retain Fermilab's accelerator expertise.

It should be pointed out that the 8 GeV RCS in this hybrid design is a different and improved version from the previous design in Ref. [2] thanks to a number of new boundary conditions.

- The antiproton source will cease operation after the Tevatron collider experiment is completed in 2010 or 2011. This makes the tunnel available to house a new RCS. Consequently, a new lattice with a triangular shape has been designed [3]. It has a larger circumference than the previous design (505 m instead of 474 m), which provides more space for straight sections and correctors.
- The increased injection energy (2 GeV vs. 600 MeV in the previous design) makes the magnet aperture smaller and further reduces the RCS construction cost.

- The Recycler will be used to accumulate beams from the RCS so that the Main Injector cycle time is shortened.

# MACHINE LAYOUT AND MAJOR DESIGN PARAMETERS

This hybrid design consists of a 2 GeV superconducting $H^-$ linac and an 8 GeV rapid cycling proton synchrotron (RCS). The latter is located in the existing Debuncher enclosure. Figure 1 is a layout of this design. Table 1 lists the major design parameters.

Table 1: Major Parameters

| | |
|---|---|
| **Linac** (operating at 15 Hz) | |
| Kinetic energy (GeV) | 2 |
| Peak current (mA) | 10 |
| Pulse length (µs) | 430 |
| $H^-$ per pulse | $2.7 \times 10^{13}$ |
| Average beam current (µA) | 65 |
| Beam power (MW) | 0.13 |
| **RCS** (operating at 15 Hz) | |
| Circumference (m) | 505.3 |
| Extr kinetic energy (GeV) | 8 |
| Protons per bunch | $3 \times 10^{11}$ |
| Number of bunches | 90 |
| Protons per cycle | $2.7 \times 10^{13}$ |
| Protons per second | $4 \times 10^{14}$ |
| Norm. trans emit (mm-mrad) | $40\pi$ |
| Longitudinal emit (eV-s) | 0.2 |
| RF frequency (MHz) | 53 |
| Average beam current (µA) | 65 |
| Beam power (MW) | 0.5 |

# LINAC

The linac is similar to the first 2 GeV portion of the present 8 GeV SRF linac design. The R&D work on the 8 GeV SRF linac can be directly applied to this new design. There are, however, main differences in the repetition rate, beam current and pulse length as shown in Table 2. The ILC adopts 5 Hz repetition rate because of the Damping Rings, which need enough damping time to reduce the beam emittance to the required value. But this is unnecessary for Project-X. When we raise the repetition rate from 5 to 15 Hz to match the RCS rep rate, the beam current is reduced by one half and pulse length reduced by two thirds because the RCS is much smaller than the Recycler. Consequently the peak power requirement of the RF source (klystrons and modulators) as well as cryogenic system is also reduced by half. This would result in significant savings in the linac cost.

\*Work supported by Fermi Research Alliance, LLC under Contract No. DE-AC02-07CH11359 with the United States Department of Energy.
#chou@fnal.gov

Table 2: Comparison of Linac Parameters

| Parameter | 8 GeV Linac | 2 GeV Linac |
|---|---|---|
| Repetition rate | 5 Hz | 15 Hz |
| Beam current | 20 mA | 10 mA |
| Pulse length | 1.25 ms | 0.43 ms |
| Peak RF power per cavity (25 MV/cavity) | 0.5 MW | 0.25 MW |
| Peak klystron power (16 cavities/klystron) | 10 MW | 5 MW |

The superconducting RF components of the 2 GeV linac are almost identical to their counterparts in the 8 GeV SRF linac. The linac uses 56 S-ILC cavities in 7 cryomodules and 32 ILC cavities in 4 Type-4 cryomodules as listed in Table 3. For the klystrons, one needs a total of three (10 MW each) or six (5 MW each).

Table 3: SCRF Components of the 2 GeV Linac

| Parameter | $\beta = 0.81$ Section | $\beta = 1$ Section |
|---|---|---|
| Energy range | 0.42 – 1.3 GeV | 1.3 – 2 GeV |
| Cavity Type | S-ILC | ILC |
| # Cavities | 56 | 32 |
| # Cryomodules | 7 | 4 (Type 4) |
| # Klystrons | 2 (10 MW each) | 1 (10 MW) |

# SYNCHROTRON

The central piece of the synchrotron is a triangular lattice [3]. It has the same shape and size as the Debuncher and can readily fit to the Debuncher footprint. The triangular lattice is transition-free thanks to its high gamma-t (18.6). This is a simple doublet lattice and uses only one type of bends and one type of quads. The scheme of missing magnet in mid-cell provides zero-dispersion straight sections without dispersion suppressor. The technical systems (RF, magnet, vacuum, etc.) are revised versions of the previous design in Ref. [2]. The main parameters are listed in Table 4.

# TRANSPORT LINES

A 2 GeV $H^-$ particle transport line has no blackbody radiation stripping problem because the Doppler shift of photon energy is much smaller than that at 8 GeV [4]. Therefore, there is no need for a cryogenic beam screen. The bending magnets can be four times stronger than that in the 8 GeV design. These lead to a shorter and simpler $H^-$ beam line. The transport of protons can reuse a portion of the existing pbar and MI-8 line as shown in Figure 1.

# COST COMPARISON

A detailed cost comparison between this hybrid design and an 8 GeV SRF linac is under way and will be published in a Fermilab TM note. A preliminary rough estimate shows the hybrid design could lower the total project cost as much as 40-50 %.

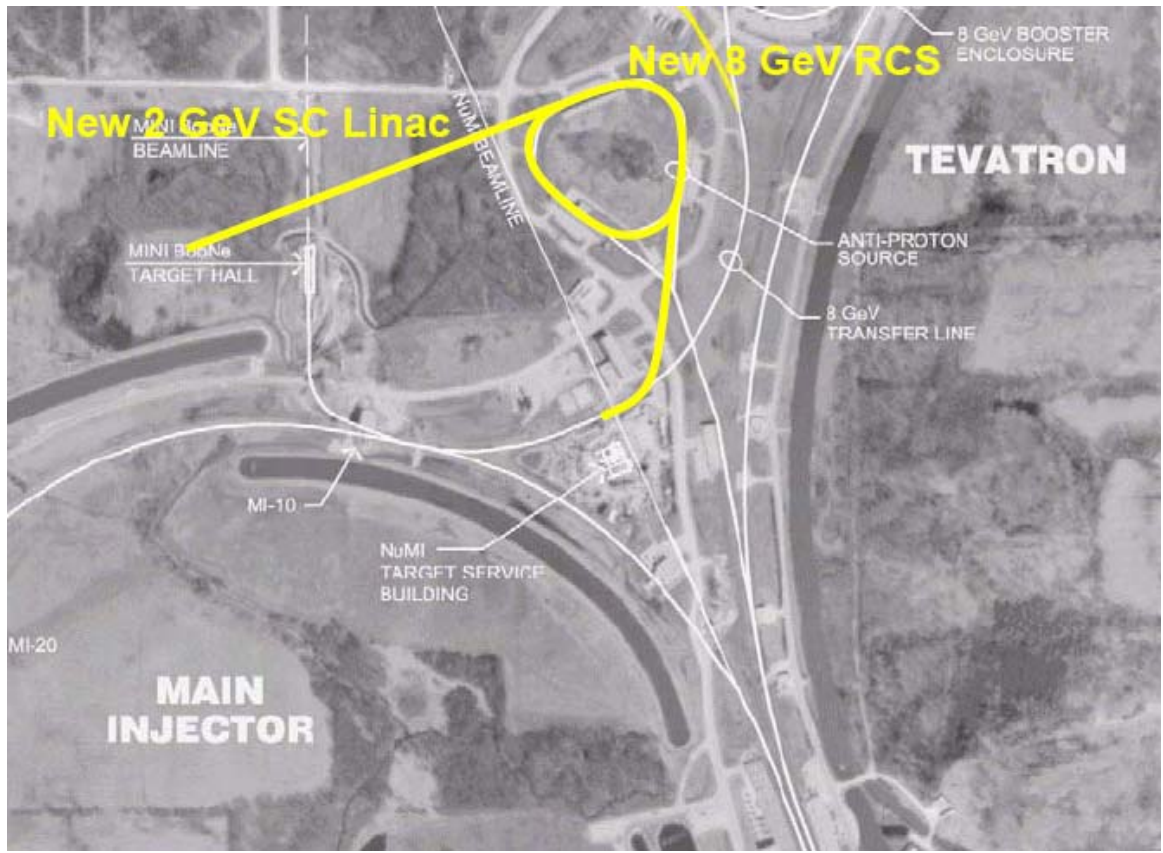

Figure 1: Layout of a 2 GeV linac and an 8 GeV RCS.

Table 4: Synchrotron Parameters

| | |
|---|---|
| Circumference (m) | 505.3 |
| Super-periodicity | 3 |
| Number of straight sections | 3 |
| Length of each arc (m) | 126.3 |
| Length of each straight section (m) | 42.1 |
| Injection kinetic energy (GeV) | 2 |
| Extraction kinetic energy (GeV) | 8 |
| Injection dipole field (T) | 0.37 |
| Peak dipole field (T) | 1.5 |
| Bending radius (m) | 19.86 |
| Peak quad gradient (T/m) | 10.26 |
| Good field region | 3 in × 5 in |
| Number of dipoles (5.2 m each) | 24 |
| Number of quads (1.238 m each) | 96 |
|    In the arcs | 72 |
|    In the straight sections | 24 |
| Max $\beta_x$, $\beta_y$ (m) | 15.6, 16.2 |
| Min $\beta_x$, $\beta_y$ (m) | 4.2, 3.8 |
| Max $D_x$ in the arcs (m) | 2.4 |
| Dispersion in the straight | 0 |
| Transition $\gamma_t$ | 18.6 |
| Horizontal, vertical tune $\nu_x$, $\nu_y$ | 11.98, 12.21 |
| Natural chromaticity $\xi_x$, $\xi_y$ | −15.1, −15.5 |
| Revolution time at inj./extr ($\mu$s) | 1.8, 1.7 |
| Injection time ($\mu$s) | 430 |
| Injection turns | 240 |
| Maximum Laslett tune shift | 0.15 |
| Norm. transv. emittance $\varepsilon_N$ (mm-mrad) | |
|    Injection beam (95%) | 2.5$\pi$ |
|    Circulating beam (100%) | 40$\pi$ |
| Longitudinal emittance (95%, eV-s) | |
|    Injection beam | 0.1 |
|    Circulating beam | 0.2 |
| Extracted bunch length $\sigma_t$ (rms, ns) | 1 |
| Momentum acceptance $\Delta p/p$ | ±1% |
| Dynamic aperture | 350$\pi$ |